\documentclass[%
 reprint,
 amsmath,amssymb,
 aps,
]{revtex4-2}

\usepackage{graphicx}
\usepackage{dcolumn}
\usepackage{bm}
\usepackage[breaklinks]{hyperref}
\usepackage{braket}
\usepackage{color}
\def\tr{\operatorname{tr}}
\renewcommand{\(}{\left(}
\renewcommand{\)}{\right)}
\renewcommand{\[}{\left[}
\renewcommand{\]}{\right]}
\newcommand{\Eqref}[1]{Eq.~\eqref{#1}}

\newcommand{\figref}[1]{Fig.~(\ref{#1})}
\renewcommand{\(}{\left(}
\renewcommand{\)}{\right)}
\renewcommand{\[}{\left[}
\renewcommand{\]}{\right]}
\let\a=\alpha \let\b=\beta \let\g=\gamma \let\d=\delta 
   
     \let\r=\rho

\let\y=\psi
  \let\D=\Delta

\newcommand{\bR}{\bar{R}}
\usepackage{amsthm}
\theoremstyle{plain}

\theoremstyle{definition}

\definecolor{bcolor}{rgb}{0.1,0,1}

\begin{document}

\preprint{APS/123-QED}

\title{The Diagonal Approximation for Holographic R\'{e}nyi Entropies}

\author{Geoff Penington}
\email{geoffp@berkeley.edu}
\author{Pratik Rath}
\email{pratik\_rath@berkeley.edu}
\affiliation{Center for Theoretical Physics and Department of Physics,
University of California, Berkeley, CA 94720, USA
}%

\begin{abstract}
Recently, Ref.~\cite{Dong:2023bfy} proposed a modified cosmic brane prescription for computing the R\'{e}nyi entropy $S_\alpha$ of a holographic system in the presence of multiple extremal surfaces. This prescription was found by assuming a diagonal approximation, where the R\'{e}nyi entropy is computed after first measuring the areas of all extremal surfaces. We derive this diagonal approximation and show that it accurately computes R\'{e}nyi entropies up to $O(\log G)$ corrections. For $\alpha<1$, this allows us to derive the modified cosmic brane prescription, which differs from the original cosmic brane prescription at leading order in $G$. For $\alpha>1$, it leads to the original cosmic brane prescription without needing to assume that replica symmetry is unbroken in the bulk. 
\end{abstract}

\maketitle

\section{Introduction}

The R\'{e}nyi entropy of a reduced density matrix $\r_R$ defined as
\begin{equation}
    S_\a(\r_R)=\frac{1}{1-\a}\log \tr\(\r_R^\a\),
\end{equation}
is a useful measure of bipartite entanglement between subsystem $R$ and its complement $\bR$.
In particular, it reduces to the entanglement entropy in the $\a\to 1$ limit.

In the context of holography, the R\'{e}nyi entropy has played an important role in our understanding of the emergence of bulk spacetime from underlying entanglement in the boundary theory.
A prescription for the holographic dual for the R\'{e}nyi entropy was made in Ref.~\cite{Lewkowycz:2013nqa}. A prescription for a related quantity called the refined R\'{e}nyi entropy was made in Ref.~\cite{Dong:2016fnf}. These prescriptions will collectively be referred to as the original cosmic brane prescription.
As a special case of these prescriptions, the $\a\to1$ limit resulted in the Ryu-Takayanagi (RT) formula that relates the entanglement entropy to the area of the minimal area surface homologous to $R$ \cite{Ryu:2006bv}.

More recently, however, it has been noticed, in various calculations, that the original cosmic brane prescription appears to fail fairly generically when $\a<1$ and there are multiple extremal surfaces present in the spacetime \cite{Murthy:2019qvb,Dong:2021oad,Akers:2022max}. This led Ref.~\cite{Dong:2023bfy} to propose a modified cosmic brane prescription that agrees with the original cosmic brane prescription for $\a>1$ while providing an improved answer for $\a<1$ that matches the calculations of Refs.~\cite{Murthy:2019qvb,Dong:2021oad,Akers:2022max}. 

The modified cosmic brane prescription of Ref.~\cite{Dong:2023bfy} was based on assuming a diagonal approximation for the R\'{e}nyi entropy in the basis of fixed-area states, which we review below. This prescription has since been applied to obtain a holographic dual of entanglement negativity \cite{Dong:2024gud}. A diagonal approximation of this form has since also been utilized by Refs.~\cite{Chen:2024lji,Chen:2024ysb} to obtain results for R\'{e}nyi entropy in 2D CFT. 

In this short note, we derive this diagonal approximation from first principles. Consequently, we are able to derive the modified cosmic brane prescription from the holographic prescription for the R\'{e}nyi entropies of fixed-area states \cite{Akers:2018fow,Dong:2018seb,Dong:2019piw}.

\section{Diagonal Approximation}
\label{sec:diag}

\begin{figure}[t]
\includegraphics[scale=0.5]{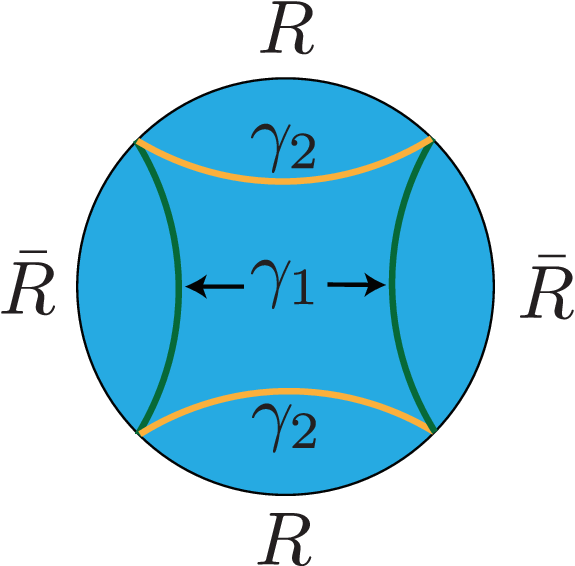}
\caption{Cauchy slice of a geometry with subregion $R$ chosen to be two disjoint intervals. There are two candidate RT surfaces $\g_1$ (green) and $\g_2$ (orange).}\label{fig:conv}
\end{figure}

Consider a general holographic state $\ket{\y}$ such that subregion $R$ has two candidate RT surfaces $\g_1$ and $\g_2$, e.g., see \figref{fig:conv}. The state $\ket{\y}$ can be decomposed into a basis of fixed-area states $\ket{A_1,A_2}$ where $A_i$ specifies the area of surface $\g_i$ \cite{Akers:2018fow,Dong:2018seb,Dong:2019piw}. The decomposition of $\ket{\y}$ takes the form \cite{Marolf:2020vsi,Akers:2020pmf}:
\begin{equation}\label{eq:sup}
    \ket{\y} = \sum_{A_1,A_2} \sqrt{p\(A_1,A_2\)} \ket{A_1,A_2},
\end{equation}
where $p(A_1,A_2)$ is the probability distribution over the two areas in state $\ket{\y}$. Here, we are considering the area to be fixed to a precision $\Delta\sim G^\b$ with $\b\geq 1$ and thus, the sum in \Eqref{eq:sup} is discretized into windows of size $\Delta$. The distribution $p\(A_1,A_2\)$ can be computed using the gravitational path integral as
\begin{equation}\label{eq:path}
    p(A_1,A_2) = \frac{\bra{\y}\Pi_{A_1,A_2}\ket{\y}}{\langle\y|\y\rangle} = \exp\(I\[g_\y\]-I\[g_{A_1,A_2}\]\),
\end{equation}
where $\Pi_{A_1,A_2}$ is a projector onto definite values of the areas of the two surfaces, $I$ is the bulk gravitational action, $g_\y$ is the smooth Euclidean geometry computing the norm of the state $\ket{\y}$ and $g_{A_1,A_2}$ is the corresponding fixed-area geometry, which will in general have conical singularities at the surfaces $\gamma_1$ and $\gamma_2$.

Given this decomposition, we can then obtain the reduced density matrix
\begin{equation}\label{eq:red}
    \r_R = \sum_{A_{1,2},A_{1,2}'} \sqrt{p\(A_1,A_2\)p\(A_1',A_2'\)}\tr_{\bR}\(\ket{A_1,A_2}\bra{A_1',A_2'}\).
\end{equation}
Using \Eqref{eq:red} to compute the R\'{e}nyi entropy, we obtain two kinds of terms: diagonal terms with $A_1=A_1'$ and $A_2=A_2'$, as well as off-diagonal terms. Ref.~\cite{Dong:2023bfy} argued that at integer $\a$, only diagonal terms contribute to replica-symmetric bulk configurations. The assumption of bulk replica symmetry used in Ref.~\cite{Lewkowycz:2013nqa} thus ignores all effects from off-diagonal terms. This means that the R\'{e}nyi entropy $S_\a(\rho_R)$ is indistinguishable from the R\'{e}nyi entropy $S_\a(\tilde \rho_R)$ of the state
\begin{align}
    \tilde \r_R = \sum_{A_{1,2}} p\(A_1,A_2\)\tr_{\bR}\(\ket{A_1,A_2}\bra{A_1,A_2}\).
\end{align}

In holography, the probability distribution found using \Eqref{eq:path} is a sharply peaked, continuous distribution which can be written in the form
\begin{equation}\label{eq:prob}
    p(A_1,A_2) \equiv \exp\[-\frac{f\(A_1,A_2\)}{G}\].
\end{equation}
Similarly the R\'{e}nyi entropies of $\r_R\(A_1,A_2\)$, the reduced density matrix on $R$ in a fixed-area state, are given by
\begin{equation}\label{eq:ent}
    \tr\(\r_R\(A_1,A_2\)^\a\) = \D^{1-\a} \exp\[-(\a-1)\min_{i=1,2} \frac{A_i}{4G}\],
\end{equation}
where $\Delta$ is the width of the fixed-area state. Combining \Eqref{eq:prob} and \Eqref{eq:ent}, and applying a saddle point approximation, we have
\begin{equation}\label{eq:max}
    \tr\(\tilde\rho_R^\a\)\sim \max_{A_1,A_2} \exp\[-\frac{1}{G}\(f\(A_1,A_2\)+(n-1)\min_{i=1,2} \frac{A_i}{4}\)\].
\end{equation}
Here, and throughout this paper, the notation $\sim$ means that two sides are equal up to factors that are polynomial in $G$. This includes, in particular, factors that are polynomial in $\D$. The R\'{e}nyi entropy $S_\a(\tilde\rho_R)$ is therefore given, up to $O(\log G)$ corrections, by a logarithm of the right-hand side of \Eqref{eq:max} divided by $(1-\a)$. The modified cosmic brane prescription of Ref.~\cite{Dong:2023bfy} proposes that this is also the R\'{e}nyi entropy $S_\a(\rho_R)$ of the original density matrix $\rho_R$, up to (potentially different) $O(\log G)$ corrections.

A diagonal approximation, where the state $\rho_R$ was replaced by $\tilde\rho_R$, was initially proposed for computing von Neumann entropies by Ref.~\cite{Marolf:2020vsi} in order to argue for $O\(\frac{1}{\sqrt{G}}\)$ corrections to the entanglement entropy near phase transitions. This approximation was later derived, up to $O(\log G)$ corrections, by Ref.~\cite{Akers:2020pmf}. 

The derivation in Ref.~\cite{Akers:2020pmf} only applies for computations of von Neumann entropies. However, we will now use similar ideas to derive the diagonal approximation for general R\'{e}nyi entropies and justify the modified cosmic brane prescription.

\section{Derivation}
\label{sec:arg}
Our argument for the diagonal approximation will closely follow the analogous argument of Ref.~\cite{Akers:2020pmf} for the entanglement entropy. The general idea will be to use entanglement wedge reconstruction to argue that subregion $R$ can always measure $A_2$ and similarly, $A_1$ can always be measured by $\bR$. Using this, we will infer that the relevant density matrices are mixtures of fixed-area density matrices. Finally, we will obtain the diagonal approximation by sandwiching the R\'{e}nyi entropy of a mixture using various inequalities. We now separately consider the cases $\a<1$ and $\a>1$.

\subsection{$\a<1$}

First, consider taking only a superposition over different values of $A_1$ for a fixed value of $A_2$. As stated above, since $A_1$ is always measurable by $\bR$, the reduced density matrix on $R$ becomes an incoherent mixture, i.e.,
\begin{align}\label{eq:mix1}
    \r_R(A_2) &= \Pi_{A_2} \rho_R \Pi_{A_2}
    \\&= \sum_{A_1} p(A_1|A_2)\tr_{\bR}\(\ket{A_1,A_2}\bra{A_1,A_2}\),\nonumber
\end{align}
with no off-diagonal terms where $A_1' \neq A_1$. Here, $\Pi_{A_2}$ is a projector onto states where $\gamma_2$ has area $A_2$ and $p(A_1|A_2)=\frac{p(A_1,A_2)}{p(A_2)}$ is the conditional probability that $\gamma_1$ has area $A_1$, assuming $\gamma_2$ has area $A_2$. \Eqref{eq:mix1} is an exact statement at the level of saddles in the gravitational path integral. In a full theory of quantum gravity, the area $A_2$ is presumably only approximately measurable up to nonperturbative corrections, but the error introduced by this should not affect our results at the order of interest.

Now, for any mixture, we will prove the following useful inequalities to upper and lower bound the R\'{e}nyi entropy. First, we note that
\begin{equation}
    \sum_i p_i \r_i \geq \max_i p_i\r_i,
\end{equation}
which holds as an operator inequality since subtracting the right-hand side (RHS) from the left-hand side (LHS) still leaves us with a positive sum of positive operators. Then we note that $X^\a$ for $\a\in\[0,1\]$ is an operator monotone, i.e.,
\begin{equation}\label{eq:hein}
    X\geq Y \implies X^\a \geq Y^\a,
\end{equation}
where $X\geq Y$ means that $X-Y$ is a positive semidefinite operator. \Eqref{eq:hein} is known as the Lowner-Heinz inequality. Using this for \Eqref{eq:mix1} we obtain,
\begin{align}\label{eq:low1}
    \tr\(\r_R(A_2)^\a\) &\geq \max_{A_1} p\(A_1|A_2\)^\a \tr\(\rho_R(A_1,A_2)^\a\).
\end{align}
This provides a lower bound  for the R\'{e}nyi entropy. 

In order to obtain an upper bound, we use the following majorization result:
\begin{equation}\label{eq:maj}
    \oplus_{i} p_i \rho_i \prec \sum_i p_i \r_i,
\end{equation}
which means that after sorting the eigenvalues in decreasing order, the sum of the first $k$ eigenvalues of the LHS is smaller than (or equal to) that of the RHS. For more details, we refer the reader to Ref.~\cite{nielsen}. Then, we can use the fact that $\tr\(X^\a\)$ for $\a\in\(0,1\)$ is a Schur concave function, i.e.,
\begin{equation}
    X \prec Y \implies \tr\(X^\a\) \geq \tr\(Y^\a\),
\end{equation}
which applied to \Eqref{eq:mix1} gives us
\begin{equation}\label{eq:up1}
    \tr\(\r_R(A_2)^\a\) \leq \sum_{A_1} p\(A_1|A_2\)^\a \tr\(\rho_R(A_1,A_2)^\a\).
\end{equation}
This is an upper bound for the R\'{e}nyi entropy.

Now we note that the upper and lower bounds both involve the same sharply peaked function which has the form
\begin{equation}\label{eq:def}
    p\(A_1|A_2\)^\a \tr\(\rho_R(A_1,A_2)^\a\) \equiv \exp\[-\frac{g_{\a,A_2}(A_1)}{G}\].
\end{equation}
While the lower bound of \Eqref{eq:low1} is given by the peak value, the upper bound of \Eqref{eq:up1} receives contributions from the sum near the peak value. \Eqref{eq:def} emphasizes that the function has a width of $O(\sqrt{G})$ near its peak.
Thus, in the semiclassical approximation, the upper and lower bounds agree up to a factor of $O(G^{\b-\frac{1}{2}})$. 
Such factors will only affect the R\'{e}nyi entropy at $O(\log G)$. We therefore have
\begin{equation}\label{eq:first}
    \tr\(\rho_R(A_2)^\a\)\sim \max_{A_1} p\(A_1|A_2\)^\a \exp\[-(\a-1)\min_i \frac{A_i}{4G}\].
\end{equation}

We can now add in a superposition over different values of $A_2$. To do so, we consider the complementary state $\rho_{\bR}$ on $\bR$ which satisfies $S_\a\(\rho_{\bR}\) = S_\a(\rho_R)$. Since $A_2$ can always be measured by $R$, $\rho_{\bR}$ decomposes into an incoherent mixture
\begin{equation}
    \r_{\bR} = \sum_{A_2} p(A_2) \r_{\bR}(A_2)
\end{equation}
We can therefore repeat the same argument that we used above, and apply the inequalities of \Eqref{eq:low1} and \Eqref{eq:up1} to $\r_{\bR}$. This leads to
\begin{equation}
    \tr\(\rho_{\bR}^\a\)\sim \max_{A_2} p\(A_2\)^\a \tr\(\rho_{\bR}(A_2)^\a\).
\end{equation}
Finally using the fact that $\tr \(\r_R^\a\) = \tr \(\r_{\bR}^\a\)$ and \Eqref{eq:first}, we obtain
\begin{align}
    \tr\(\rho_{R}^\a\)&\sim \max_{A_2} p\(A_2\)^\a \tr\(\rho_{\bR}(A_2)^\a\)\\
    &\sim \max_{A_1,A_2} p(A_1,A_2)^\a \exp\[-(\a-1)\min_i\frac{A_i}{4G}\],\label{eq:final}
\end{align}
where we remind the reader that $p(A_1,A_2)=p(A_2)p(A_1|A_2)$. \Eqref{eq:final} is precisely the diagonal approximation \Eqref{eq:max}, which we have now proved for $\a<1$.

\subsection{$\a>1$}

The argument for $\a>1$ is quite similar except the relevant inequalities bounding the R\'{e}nyi entropy are different. Thus, we will only focus on the upper and lower bounds on the R\'{e}nyi entropy of a mixture of density matrices. With this replacement, the argument for the diagonal approximation is identical to the case of $\a<1$.

Consider a mixture of density matrices $\r=\sum_i p_i \r_i$. For an upper bound on $\tr\(\r^\a\)$ for $\a>1$, we can use the fact that the Schatten $p$-norm defined as
\begin{equation}
    ||A||_p = \(\tr\(A^p\)\)^{1/p}
\end{equation}
satisfies the triangle inequality. Thus, we have
\begin{equation}\label{eq:tri}
    ||\sum_i A_i||_p \leq \sum_i ||A_i||_p
\end{equation}
Applying \Eqref{eq:tri} in the case $p=\a$ to $\r$, the mixture of density matrices, this implies that
\begin{equation}\label{eq:up2}
    \tr\(\r^\a\)\leq \(\sum_i p_i \(\tr\(\r^\a\)\)^{1/\a}\)^\a
\end{equation}
for all $\a\geq 1$. This inequality is an improved version of the inequality applied in Ref.~\cite{Dong:2023xxe}. 

On the other hand, to obtain the lower bound we can again use the majorization result \Eqref{eq:maj} along with the fact that $\tr\(X^\a\)$ is Schur-convex for $\a>1$ to obtain the inequality
\begin{equation}\label{eq:low2}
    \tr\(\r^\a\)\geq \sum_i p_i^\a \tr(\r_i^\a).
\end{equation}

In our case of interest, we work in the semiclassical approximation and note that the upper and lower bounds both involve the same function given in \Eqref{eq:def}. Since we have chosen $\D$ to be parametrically small in $G$, we can approximate the sums by integrals. Denoting the upper bound by $U$ and the lower bound by $L$, we have
\begin{align}
    L &= \int dx \exp\[-\frac{g_\a(x)}{G}\]\\
    U &= \(\int dx \exp\[-\frac{g_\a(x)}{\a \,G}\]\)^\a.
\end{align}
Evaluating these integrals in the saddle point approximation, it is clear that $U$ and $L$ agree up to $O(G^\d)$ factors. This means that the diagonal approximation holds for the R\'{e}nyi entropy up to errors of $O(\log G)$.

\section{Summary and Discussion}

We derived the diagonal approximation that was used by Ref.~\cite{Dong:2023bfy} to obtain the modified cosmic brane prescription. Since the modified cosmic brane prescription generically corrects the original cosmic brane prescription at leading order in $G$ when $\a<1$, our argument puts those violations of the original cosmic brane prescription on a much stronger footing. 

The original cosmic brane proposal of Ref.~\cite{Lewkowycz:2013nqa} was found by assuming the dominant bulk saddle used to compute a R\'{e}nyi entropy is replica symmetric and hence can be captured by a quotiented geometry involving only a single replica. For integer $\a >1$, this assumption seems fairly robust. However, to derive the results of Ref.~\cite{Lewkowycz:2013nqa} this approximation needs to continue to be valid after analytic continuation to noninteger $\a$. The divergence of the modified and original cosmic brane prescriptions show that this analytic continuation can generically fail once $\a < 1$.

We can instead use the diagonal approximation to argue for the original cosmic brane proposal for $\a>1$, assuming the holographic R\'{e}nyi entropy prescription for fixed-area states~\cite{Akers:2018fow,Dong:2018seb,Dong:2019piw}. This is because, when $\a>1$, the minimization over surfaces $\gamma_1$, $\gamma_2$ in \Eqref{eq:max} becomes a maximization when taken outside the exponential. As explained in Ref.~\cite{Dong:2023bfy}, it can then be exchanged with the maximization over $A_1$, $A_2$ and one recovers the original cosmic brane prescription.

The prescription for fixed-area state holographic R\'{e}nyi entropies itself can be derived from an approximation to the gravitational path integral where we sum over all gravitational saddles, including replica-symmetry breaking saddles, that can be found by gluing copies of the original geometry together, with twist operators, permuting different replicas, inserted at $\gamma_1$ and $\gamma_2$ \cite{Penington:2019kki}. The approximations required for that argument are much weaker than the assumption of bulk replica symmetry used in Ref.~\cite{Lewkowycz:2013nqa}.  However, for any $\a > 1$, the results of this paper allow us to derive all the results of Ref.~\cite{Lewkowycz:2013nqa} from those weaker assumptions. 

The assumption of bulk replica symmetry has also been used to derive subleading corrections to the gravitational entropy for bulk matter fields \cite{Faulkner:2013ana,Engelhardt:2014gca}. For $\a > 1$, and for $\a < 1$ whenever \eqref{eq:max} is unchanged if we exchange the order of maximisation and minimisation (i.e. the original cosmic brane prescription is valid), we strongly expect those corrections to be correct. However, our current proof method is insufficient to show this because we only have control over the R\'{e}nyi entropies up to $O(\log G)$ corrections. We leave further analysis of this issue to future work. When $\a < 1$ and the original cosmic brane proposal fails, we do not expect a simple formula for $O(1)$ corrections to the R\'{e}nyi entropy to exist. Such situations are somewhat analogous to the situations considered in Refs.~\cite{Akers:2020pmf, Akers:2023fqr} where the min- and max-entanglement wedges for a boundary region differ and no simple formula for the boundary von Neumann entropy exists.

\begin{acknowledgments}
PR would like to thank Anurag Sahay for discussions. This work was supported in part by the Berkeley Center for Theoretical Physics, by the Department of Energy, Office of Science, Office of High Energy Physics through DE-SC0019380 and  DE-FOA-0002563, by AFOSR award FA9550-22-1-0098 and by a Sloan Fellowship.
\end{acknowledgments}

\bibliography{apssamp}

\end{document}